\documentstyle[12pt]{article}
\begin{document}
\hfill{CCUTH-96-04}
\vskip 0.5cm
\begin{center}
{\large {\bf Resummation at Large $Q$ and at Small $x$}}
\vskip 1.0cm
Hsiang-nan Li
\vskip 0.5cm
Department of Physics, National Chung-Cheng University, \par
Chia-Yi, Taiwan, Republic of China
\end{center}
\vskip 2.0cm

PACS numbers: 12.38.Bx, 12.38.Cy
\vskip 2.0cm

\centerline{\bf Abstract}
\vskip 0.3cm
We propose a unified and simple viewpoint to various evolution equations 
appropriate in different kinematic regions. We show that the resummation 
technique reduces to the Altarelli-Parisi equation, if the transverse 
degrees of freedom of a parton are ignored, to the 
Balitskii-Fadin-Kuraev-Lipatov equation, if the momentum fraction of a 
parton vanishes, and to the Gribov-Levin-Ryskin equation, if the 
recombination of gluons is included. 

\newpage

Recently, we applied the resummation technique to hard QCD processes,
such as deeply inelastic scattering and Drell-Yan production, and grouped
large radiative corrections into Sudakov factors \cite{L1}. In particular,
we have demonstrated how to resum the double logarithms produced in 
radiative corrections to a quark distribution function. In this letter we 
shall show that the resummation technique can also deal with 
single-logarithm cases appropriate in different kinematic
regions. It reduces to the conventional evolution equations, such as the 
Altarelli-Parisi (AP) equation \cite{AP}, which organizes large logarithms
$\ln Q$, $Q$ being a momentum transfer, the 
Balitskii-Fadin-Kuraev-Lipatov (BFKL) equation \cite{BFKL}, which organizes
large $\ln(1/x)$, $x$ being the Bjorken variable,
and the Gribov-Levin-Ryskin equation \cite{GLR}, which organizes
large $\ln Q\ln(1/x)$. It has been shown that the resummation technique is
equivalent to the Wilson-loop formalism for the summation of soft 
logarithms in \cite{L1,L2}. Hence, we provide a unified viewpoint to all 
these methods in terms of the concept of the resummation. It will be found
that the derivation of the above equations by means of the resummation
is simpler. 

\noindent
1. The Altarelli-Parisi equation

We consider a flavor nonsinglet structure function associated with deeply 
inelastic scattering of a hadron with momentum $p^\mu=p^+\delta^{\mu +}$. 
A quark distribution funtion $\phi$ is constructed 
by factorizing collinear radiative gluons with momenta parallel to $p$ 
from the hard scattering amplitude onto an eikonal 
line in an arbitrary direction $n$. In axial gauge $n\cdot A=0$ these 
collinear gluons are decoupled from the eikonal line, and the $n$ 
dependence goes into the gluon propagator $(-i/l^2)N^{\mu\nu}(l)$ with
$N^{\mu\nu}=g^{\mu\nu}-(n^\mu l^\nu+n^\nu l^\mu)/
n\cdot l+n^2l^\mu l^\nu/(n\cdot l)^2$.

The key step in the resummation is to obtain the derivative
$p^+d\phi/dp^+$. It has been argued 
that $\phi$ depends only on the ratio $(p\cdot n)^2/n^2$ \cite{L1}. 
A chain rule relating $p^+d/dp^+$ to  $d/dn$ leads to the formula \cite{L1}
\begin{equation}
p^+\frac{d}{dp^+}\phi(x,p^+,\mu)=2{\tilde \phi}(x,p^+,\mu)\;,
\label{dph}
\end{equation}
shown in Fig.~1(a), where $x=-q^2/(2p\cdot q)$ is the Bjorken variable with 
$q$ the momentum transfer. The argument $p^+$ denotes the large
logarithms $\ln (p^+/\mu)$ appearing in $\phi$.
The function $\tilde \phi$ contains a new vertex at the outer end 
of the quark line, which is represented by a square and expressed as 
${\hat v}_\alpha=n^2v_\alpha/(v\cdot nn\cdot l)$ \cite{L1}
with $v_\alpha=\delta_{\alpha +}$ a vector along $p$. 
This vertex is obtained by applying 
$d/dn_\alpha$ to the gluon propagator $N^{\mu\nu}$.
The coefficient 2 counts the two valence quark lines. 

We show that Eq.~(\ref{dph}) reduces to the AP equation,
if the transverse degrees of freedom of a parton are ignored.
For simplicity, we concentrate on the end-point region with $x\to 1$. 
As stated in \cite{L1}, the leading regions of the loop 
momentum flowing through the new vertex are soft and hard, in which 
$\tilde \phi$ can be factorized into the convolution of the subdiagram 
containing the new vertex with the original distribution function $\phi$. 
$\tilde \phi$ is then written as
\begin{eqnarray}
{\tilde \phi}(x,p^+,\mu)=
\int_x^1 d\xi [K(x,\xi,\mu)+G(x,\xi,\mu)]\phi(\xi,p^+,\mu)\;,
\label{apeq}
\end{eqnarray}
where the function $K$, absorbing soft divergences, corresponds to
Fig.~1(b), and $G$, absorbing ultraviolet divergences, 
corresponds to Fig.~1(c). Their expressions are given by
\begin{eqnarray}
K&=&ig^2{\cal C}_F\mu^\epsilon\int
\frac{d^{4-\epsilon}l}{(2\pi)^{4-\epsilon}}
\frac{{\hat v}_\mu v_\nu}{v\cdot l}
\left[\frac{\delta(\xi-x)}{l^2}+2\pi i\delta(l^2)\delta(\xi-x-l^+/p^+)
\right]N^{\mu\nu}
\nonumber \\
& &-\delta K\;,
\label{kj}\\
G&=&ig^2{\cal C}_F\mu^\epsilon\int\frac{d^{4-\epsilon}l}
{(2\pi)^{4-\epsilon}}{\hat v}_\mu
\left(\frac{\not p+\not l}{(p+l)^2}\gamma_\nu
-\frac{v_\nu}{v\cdot l}\right)\frac{N^{\mu\nu}}{l^2}\delta(\xi-x)
-\delta G\;,
\label{gph}
\end{eqnarray}
where $\delta K$ and $\delta G$ are additive counterterms. 
A straightforward calculation gives
\begin{eqnarray}
K=\frac{\alpha_s}{\pi\xi}{\cal C}_F\left[\frac{1}{(1-x/\xi)_+}
+\ln\frac{\nu p^+}{\mu}\right]\;,
\;\;\;\;
G=-\frac{\alpha_s}{\pi\xi}{\cal C}_F\ln\frac{\nu p^+}{\mu}\;,
\end{eqnarray}
with $\nu=\sqrt{(v\cdot n)^2/|n^2|}$. 

We then treat $K$ and $G$ by renormalization group (RG) methods \cite{L1}.
The RG solution of $K+G$ is written as
\begin{eqnarray}
K(x,\xi,\mu)+G(x,\xi,\mu)&=&K(x,\xi,p^+)+G(x,\xi,p^+)-
\int_{p^+}^{p^+}\frac{d{\bar\mu}}{\bar\mu}
\lambda_K(\alpha_s({\bar\mu}))\;,
\nonumber \\
&=&\frac{\alpha_s(p^+)}{\pi\xi}{\cal C}_F\frac{1}{(1-x/\xi)_+}\;,
\label{skg}
\end{eqnarray}
where the anomalous dimension of $K$ is defined by 
$\lambda_K=\mu d\delta K/d\mu$. 
Obviously, the source of double logarithms, {\it ie.}, the integral 
cantaining $\lambda_K$, vanishes. 
If the transverse degrees of freedom are included,
the function $\delta(\xi-x-l^+/p^+)$ in Eq.~(\ref{kj}) will be replaced by
$\delta(\xi-x)\exp(i{\bf l}_T\cdot {\bf b})$, $b$ being the transverse 
distance travelled by the parton \cite{L1}. In this case the scale 
$1/b$ is substituted for the lower bound of $\bar\mu$. Double 
logarithms then exist, implying that the soft logarithms in $\phi$ do not 
cancel exactly. Therefore, the resummation technique can deal with both
the double-logarithm and single-logarithm problems.

Inserting Eq.~(\ref{skg}) into (\ref{apeq}) and solving (\ref{dph}), 
we derive
\begin{equation}
\phi(x,\mu)=\phi(x,\Lambda,\mu)+\int_\Lambda^\mu \frac{d{\bar\mu}}{\bar\mu}
\frac{\alpha_s({\bar\mu})}{\pi}{\cal C}_F
\int_x^1 \frac{d\xi}{\xi} \frac{2}{(1-x/\xi)_+}\phi(\xi,{\bar\mu},\mu)\;,
\label{sp1}
\end{equation}
with $\Lambda$ an infrared cutoff. We have defined 
$\phi(x,\mu)\equiv\phi(x,\mu,\mu)$, which does not contain large logarithms.
Differentiating Eq.~(\ref{sp1}) with respect to $\mu$, and substituting 
the RG equation $\mu d\phi/d\mu=-2\lambda_q\phi$, $\lambda_q=-\alpha_s/\pi$ 
being the quark anomalous dimension in axial gauge, we have
\begin{eqnarray}
\mu\frac{d}{d\mu}\phi(x,\mu)=
\frac{\alpha_s(\mu)}{\pi}{\cal C}_F\int_x^1 \frac{d\xi}{\xi} 
\frac{2}{(1-x/\xi)_+}\phi(\xi,\mu)-\lambda_q(\mu)\phi(x,\mu)\;,
\label{sp11}
\end{eqnarray}
where Eq.~(\ref{sp1}) has been inserted to obtain the second term on the
right-hand side of the above equation. Eq.~(\ref{sp11}) can be recast
into
\begin{eqnarray}
\mu\frac{d}{d\mu}\phi(x,\mu)=\frac{\alpha_s(\mu)}{\pi}
\int_x^1 \frac{d\xi}{\xi} P(x/\xi)\phi(\xi,\mu)\;,
\label{sp2}
\end{eqnarray}
with the function 
\begin{eqnarray}
P(x)={\cal C}_F\left[\frac{2}{(1-x)_+}+\frac{3}{2}\delta(1-x)\right]\;.
\end{eqnarray}
It is easy to identify $P$ as the splitting function 
$P_{qq}(x)={\cal C}_F[(1+x^2)/(1-x)_+$
$+(3/2)\delta(1-x)]$ in the limit $x\to 1$ \cite{AP}.
The AP equations for intermediate $x$ and for
other kinds of partons will be discussed elsewhere.

\noindent
2. The BFKL equation

We show that the master equation in the resummation reduces to 
the BFKL equation in the small $x$ region. We start with the derivative
of the gluon distribution function $F(p_L^+,p_T)$ shown in Fig.~1(a), where 
the valence partons are regarded as gluons with longitudinal momentum 
$p_L^+$ and
transverse momentum $p_T$. It is convenient to work in covariant gauge 
$\partial A=0$, under which the square vertex is replaced by an eikonal 
line in the direction $n$ along with a new vertex 
${\hat n}_\alpha=(n^2/v\cdot n)\left[v\cdot ln_\alpha/n\cdot l
-v_\alpha\right]$ on it \cite{L1}.
This vertex is represented by the symbol $\times$ in Fig.~2. It comes
from the derivative of the Feynman rules for the eikonal line with
respect to $n_\alpha$. We have shown that the resummation results
obtained in axial and covariant gauges are the same \cite{L1}.

We express the master equation as
\begin{equation}
p_L^+\frac{d}{dp_L^+}F(p_L^+,p_T)\equiv x\frac{d}{dx}F(x,p_T)
=4{\tilde F}(x,p_T)\;,
\label{df}
\end{equation}
where the variable change $p_L^+=xp^+$ has been employed.
Note that the coefficient 4, which is 
twice of the corresponding coefficient in the AP case, is due to the 
one more attachment of the radiative gluon to the quark lines in the box
diagram associated with the hard scattering.
Working on Eq.~(\ref{df}), we do not need to go into the detailed
analysis of angular ordering of radiative gluons as in the conventional
derivation of the BFKL equation. All the possible orderings have 
resided in Eq.~(\ref{df}). Therefore, it is simpler to
understand the evolution equations by means of the concept of the
resummation.

Similarly, the soft divergences of the 
subdiagram containing the new vertex are collected by Fig.~1(b), and
the ultraviolet divergences by Fig.~1(c).
Using the relation $f_{abc}t_bt_c=(i/2)Nt_a$ for the color structure,
$N=3$ being the number of colors, Fig.~1(b) gives
\begin{eqnarray}
{\tilde F}_{\rm soft}(x,p_T)&=&
\frac{i}{2}Ng^2\int\frac{d^{4}l}{(2\pi)^4}
\frac{\Gamma^{\mu\nu\lambda}{\hat n}_\nu}{(-2p_L^+v\cdot l)n\cdot l}
\left[\frac{\theta(p_T^2-l_T^2)}{l^2}F(x,p_T)\right.
\nonumber \\
& &+2\pi i\delta(l^2)F(x,|{\bf p}_T+{\bf l}_T|)\Biggr]\;,
\label{kf}
\end{eqnarray}
where the triple-gluon vertex for the vanishing loop momentum $l$ is given 
by $\Gamma^{\mu\nu\lambda}=p_L^+(g^{\mu\nu}v^{\lambda}+g^{\nu\lambda}v^{\mu}
-2g^{\lambda\mu}v^{\nu})$.
The first term in Eq.~(\ref{kf}) corresponds to the virtual gluon 
emission, where the $\theta$ function guarantees a small
$l_T$. Because of this $\theta$ function, the integral is ultraviolet
finite, and it is not necessary to introduce a renormalization scale $\mu$
here. The second term corresponds to the real gluon emission, where
$F(x,|{\bf p}_T+{\bf l}_T|)$ implies that the gluon entering the hard
scattering carries a transverse momentum ${\bf p}_T$, 
after emitting a real radiative gluon of momentum ${\bf l}_T$.

It can be shown that as $v^\lambda$ and $v^\mu$ are contracted with
a vertex in the hard scattering amplitude and a vertex in the gluon
distribution function, respectively, the resulting contributions are 
suppressed by a power $1/s$, $s=(p+q)^2$, compared to the contribution 
from the last term $v^\nu$. Absorbing the metric tensor $g^{\lambda\mu}$ 
into the gluon distribution function, and evaluating the integral
straightforwardly, Eq.~(\ref{kf}) becomes
\begin{eqnarray}
{\tilde F}_{\rm soft}(x,p_T)=
\frac{N\alpha_s}{4\pi}\int\frac{d^{2}l_T}{\pi l_T^2}
\left[\theta(p_T^2-l_T^2)F(x,p_T)-F(x,|{\bf p}_T+{\bf l}_T|)\right]\;.
\label{kf1}
\end{eqnarray}

It has been argued that when the fractional momentum $p_L^+$ of a parton 
is small, the logarithms from a loop momentum parallel to $p$
are suppressed \cite{L1}. Hence,
if we neglect the less important contribution from Fig.~1(c), and
adopt the approximation
${\tilde F}={\tilde F}_{\rm soft}$, Eq.~(\ref{df}) is written as 
\begin{equation}
\frac{dF(x,p_T)}{d\ln(1/x)}
=\frac{N\alpha_s}{\pi}\int\frac{d^{2}l_T}{\pi l_T^2}
[F(x,|{\bf p}_T+{\bf l}_T|)-\theta(p_T^2-l_T^2)F(x,p_T)]\;,
\label{sf}
\end{equation}
which is exactly the BFKL equation. It is then understood that the 
subdiagrams shown in Fig.~1(b) and 1(c) play the role of the kernel of 
the BFKL equation.

If including Fig.~1(c), we shall obtain a more
accurate equation. 
In the region with intermediate $x$, it can be shown that
the master equation (\ref{df}) 
reduces to the Ciafaloni-Catani-Fiorani-Marchesini equation 
\cite{CCFM}. We leave these subjects to a separate work.

\noindent
3. The GLR equation

In the region with both large $Q$ and small $x$, many gluons are
radiated by partons with small spatial separation among them. A new effect 
from the annihilation of two gluons into one gluon becomes important. 
Taking into account this effect, the BFKL equation is modified by a 
nonlinear term, and leads to the GLR equation \cite{GLR}, 
\begin{equation}
\frac{\partial^2xG(x,Q^2)}{\partial\ln(1/x)\partial \ln Q^2}
=\frac{N\alpha_s}{\pi}xG(x,Q^2)-\frac{\gamma\alpha_s}{Q^2R_N^2}
[xG(x,Q^2)]^2\;.
\label{glr}
\end{equation}
where the gluon density $xG$ is defined in terms of the gluon distribution
function by
\begin{equation}
xG(x,Q^2)=\int\frac{d^{2}l_T}{\pi}\theta(Q-l_T)F(x,l_T)\;.
\end{equation}
The constant $\gamma=81/16$ can be regarded as the effective coupling 
of the annihilation process, and the radius $R_N$ characterizes the 
correlation length of gluons \cite{LL}. It is obvious that the second term 
is of higher twist, which, however, becomes important as
$R_N$ is small. The minus sign in front of it indicates that
the annihilation decreases the number of gluons, and
that the rise of the gluon density at small $x$
due to the first term might saturate.

We show that the resummation technique can include the annihilation 
effect, and generates the GLR equation in a natural way. The starting 
point is still the master equation described by Fig.~1(a), 
\begin{equation}
x\frac{\partial^2xG(x,Q^2)}{\partial x\partial \ln Q^2}
=Q^2x\frac{\partial F(x,Q)}{\partial x}=4Q^2{\tilde F}(x,Q)\;.
\label{gl}
\end{equation}
However, when factorizing the subdiagram
containing the new vertex, we add one more diagram in which four gluons
coming out of the hadron attach the subdiagram as shown in Fig.~2(a).
This new diagram describes the annihilation process correctly.

The first diagram in Fig.~2(a) leads to Eq.~(\ref{kf1}), whose first
term is in fact plays the role of a soft subtraction,
such that the integral is free of infrared divergences. Hence, it is
a reasonable approximation to drop the first term, and to replace
the denominator $l_T^2$ by $Q^2$. We have
\begin{eqnarray}
{\tilde F}_{\rm soft}(x,Q)\approx
-\frac{N\alpha_s}{4\pi Q^2}\int_0^Q\frac{d^{2}l_T}{\pi}F(x,l_T)
=-\frac{N\alpha_s}{4\pi Q^2}xG(x,Q^2)\;,
\label{g1}
\end{eqnarray}
which gives the linear part of the GLR equation.

We then consider the second diagram in Fig.~2(a), whose lowest-order
diagram along with its soft subtraction, are shown in Fig.~2(b).
The color structure of Fig.~2(b) is given by
$f_{dhg}f_{cge}f_{bef}t_ht_f$. Assuming 
that the two gluons, labeled by the color indices $c$ and $d$, 
form a color-singlet pair, we extract the color factor ${\cal C}=
(-i/2)C_A^2/(N^2-1)$. With the same analysis of the contributions 
from each term of the triple-gluon vertices as employed in 
the BFKL case, the first diagram in Fig.~2(b) leads to
\begin{eqnarray}
{\tilde F}_{\rm soft}^{(2)}&=&
\frac{{\cal C}g^4}{(2\pi)^4}\int d^4 l
\Gamma^{\nu\sigma\varepsilon}\Gamma^{\lambda\xi\sigma}\Gamma^{\delta\zeta\xi}
\frac{{\hat n}_\varepsilon 2\pi i\delta((p_L^+v+l)^2)}{n\cdot l (l^2)^2
(-2p_L^+v\cdot l)}\int d^2 l'_T F^{(2)}(x,l_T,l^{'}_T)
\nonumber \\
&=&\frac{C_A^2}{N^2-1}\frac{g^4}{(2\pi)^3}\int d^4 l
\frac{n^2\delta((p_L^+v+l)^2)}{(n\cdot l)^2 (l^2)^2}
\int d^2 l'_TF^{(2)}(x,l_T,l^{'}_T)\;.
\label{gi}
\end{eqnarray}
To arrive at the second expression, we have absorbed the relevant metric 
tensors into the gluon distribution function
$F^{(2)}(x,l_T^2,l^{'2}_T)$, which describes the
probability of finding two gluons with equal
momentum fraction $x$ but different transverse momenta $l_T$ and $l'_T$. 

Integrating over $l^-$ and then $l^+$, Eq.~(\ref{gi}) becomes
\begin{equation}
{\tilde F}_{\rm soft}^{(2)}=
\frac{C_A^2}{N^2-1}\frac{\alpha_s^2}{\pi}\int\frac{d^2 l_T}{l_T^4}
\int d^2 l'_TF^{(2)}(x,l_T,l^{'}_T)\;,
\label{gi2}
\end{equation}
in which the vector $n$ has been chosen to be $(1,-1,{\bf 0})$ \cite{L1}. 
The contribution from the second diagram can be included simply by
substituting $Q^4$ for the denominator $l_T^4$. Eq.~(\ref{gi2})
is then cast into 
\begin{equation}
{\tilde F}_{\rm soft}^{(2)}=
\frac{\pi^3}{N^2-1}\left(\frac{\alpha_s C_A}{\pi}\right)^2\frac{1}{Q^4}
x^2G^{(2)}(x,Q^2)\;,
\label{gi3}
\end{equation}
with the definition
\begin{equation}
x^2G^{(2)}(x,Q^2)\equiv \int_0^Q\frac{d^2 l_T}{\pi}
\frac{d^2 l'_T}{\pi}F^{(2)}(x,l_T,l^{'}_T)\;.
\end{equation}

Substituting ${\tilde F}={\tilde F}_{\rm soft}+{\tilde F}_{\rm soft}^{(2)}$
into Eq.~(\ref{gl}), we obtain
\begin{equation}
\frac{\partial^2xG(x,Q^2)}{\partial\ln(1/x)\partial \ln Q^2}
=\frac{N\alpha_s}{\pi}xG(x,Q^2)-
\frac{4\pi^3}{N^2-1}\left(\frac{\alpha_s C_A}{\pi}\right)^2\frac{1}{Q^2}
x^2G^{(2)}(x,Q^2)\;,
\label{g4}
\end{equation}
which is exactly the same as the corresponding formula in \cite{MQ}.
Naively employing the relation $G^{(2)}=(3/2)R_NnG^2$ \cite{MQ} for a very
loosely bound necleus, where 
$n=(4\pi R_N^3/3)^{-1}$ is the nuclear number density,
we derive the GLR equation (\ref{glr}). Obviously,
the number of diagrams involved in the resummation technique
is much fewer and the calculation performed here is simpler
than in \cite{MQ}.

In conclusion, the master equation of the resummation technique
relates the derivative of the 
distribution function to a new function involving a new vertex. The 
summation of various large logarithms is embedded in this new function 
without resort to the complicated diagrammatic analysis. 
When expressing the new function
as a factorization formula, the subdiagram containing the new vertex
is exactly the kernel associated with the evolution equation. With this 
work, we are sure that the simpler resummation technique is applicable
to a large class of QCD processes. These applications will be published
elsewhere.

I thank J. Qiu for a useful discussion. This work was supported by
National Science Council of R.O.C. under Grant No. 
NSC-85-2112-M-194-009.

\newpage
\centerline{\large \bf Figure Captions}
\vskip 0.5cm

\noindent
{\bf Fig. 1.} (a) The derivative $p^+d\phi/dp^+$ in axial gauge.
(b) The $O(\alpha_s)$ function $K$. 
(c) The $O(\alpha_s)$ function $G$. 
\vskip 0.5cm

\noindent
{\bf Fig. 2.} (a) The contributions to $\tilde F$ including the 
annihilation effect in covariant gauge. (b) The soft structure of the 
$O(\alpha_s)$ subdiagram for the first diagram of (a).
\vskip 0.5cm

\end{document}